# Electric Field Models of Transcranial Magnetic Stimulation Coils with Arbitrary Geometries: Reconstruction from Incomplete Magnetic Field Measurements

Kristoffer H. Madsen[1,2], Maria Drakaki[1,3,4], Axel Thielscher[1,3]

[1]Danish Research Centre for Magnetic Resonance, Centre for Functional and Diagnostic Imaging and Research, Copenhagen University Hospital - Amager and Hvidovre, Copenhagen, Denmark

[2]Section for Cognitive Systems, Department of Applied Mathematics and Computer Science, Technical University of Denmark, Kgs. Lyngby, Denmark

[3]Magnetic Resonance Section, Department of Health Technology, Technical University of Denmark, Kgs. Lyngby, Denmark

[4]MagVenture A/S, Farum, Denmark

Correspondence by: Kristoffer H. Madsen
Email: kristofferm@drcmr.dk
Ph: (+45) 386 23407

## Abstract

**Background**:
Calculation of the electric field induced by transcranial magnetic stimulation (TMS) in the brain requires accurate models of the stimulation coils. Reconstructing models from the measured magnetic fields of coils so far worked only for flat coil geometries and required data of their full magnetic field distribution.

**Objective**:
To reconstruct models of coils with arbitrary winding geometries from spatially incomplete magnetic field measurements.

**Methods**:
Dipole approximation via minimum norm estimation with a cross validation procedure simultaneously assessing predictability and reproducibility of the field approximation. The methods were validated on both simulated and acquired magnetic flux density data. Furthermore, reconstruction of coil models from sparsely sampled data is investigated and a procedure for obtaining sparse dipole coil models based on orthogonal matching pursuit is suggested.

**Results**:
Given that measured data from regions around the coil are available, the magnetic vector potential can accurately be represented by fitted dipole models even from sparsely sampled data. Considerable sparsification of dipole models is possible while retaining accurate representation of the field also close to the coil.

**Conclusion**:
Our versatile approach removes an important hurdle for constructing coil models from measurement data and enables a flexible trade-off between the accuracy and computational efficiency of the reconstructed dipole models.

## Highlights
- Accurate dipole coil models reconstructed from incomplete magnetic field measurements
- Reconstruction is feasible for TMS coils with arbitrary winding geometries
- Flexible tradeoff between accuracy and computational efficiency by model sparsification

## Introduction

Realistic numerical calculations of the electric field induced by TMS in the brain require accurate theoretical models of the stimulation coils [1,2]. In the past, methods and automated approaches have been developed to measure the magnetic flux density of TMS coils, with the aim to obtain good theoretical models also when relevant technical parameters such as the internal winding geometries, the cable diameters and their exact position in the coil casing are not or only partially known [3–6].

A key challenge of these approaches is the reconstruction of the magnetic vector potential from the measured magnetic flux density, as knowledge of the vector potential in the complete head volume is required for the numerical field calculations [7,8]. Our previous reconstruction method [5] achieved accurate reconstructions, but required dense measurements in an extended volume around the coil (for arbitrary winding geometries) or from one half-space (for flat coils with mirrored magnetic fields in the unmeasured half). This made the practical data acquisition demanding and unfeasible for a regular use.

Here, we describe a novel reconstruction approach that generates accurate vector potential distributions for arbitrary winding geometries and that still works when the flux density around the TMS coil is incompletely sampled. We show an example of a sparse sampling strategy that leads to accurate and robust reconstructions for flux density data measured in only a few two-dimensional planes at varying distances to the coil. Finally, by controlling the number of magnetic dipoles used for constructing the theoretical coil model, we show that the method also enables a flexible trade-off between the accuracy and computational efficiency of the coil model when used in the numerical field calculations.

## Material and Methods

We consider the magnetic vector potential ($A$) and magnetic flux density ($B$) at position $r$ due to a collection of magnetic dipoles in free space indexed by $i$ at positions $r_i$ with the magnetic dipole moments $m_i$

$$A(r) = \frac{\mu_0}{4\pi} \sum_i \frac{m_i \times (r - r_i)}{|r - r_i|^3}$$

$$B(r) = \nabla \times A(r) = \frac{\mu_0}{4\pi} \sum_i \frac{1}{|r - r_i|^5} \left(3(r - r_i)(m_i \cdot (r - r_i)) - m_i |r - r_i|^2\right),$$

where $\nabla$ is the differential vector operator and $\times$ is the curl vector operator. By splitting coordinates and dipole moments into Cartesian coordinate parts we arrive at the following linear relation between the dipole moments and the magnetic flux density

$$B_x(\bm{r}) = \frac{\mu_0}{4\pi}\sum_i \frac{1}{|\bm{r}-\bm{r}_i|^5}\left(3(r_x - r_{i,x})^2 - |\bm{r}-\bm{r}_i|^2\right)m_{i,x}$$
$$+ \frac{1}{|\bm{r}-\bm{r}_i|^5}\left(3(r_x - r_{i,x})(r_y - r_{i,y}) - |\bm{r}-\bm{r}_i|^2\right)m_{i,y}$$
$$+ \frac{1}{|\bm{r}-\bm{r}_i|^5}\left(3(r_x - r_{i,x})(r_z - r_{i,z}) - |\bm{r}-\bm{r}_i|^2\right)m_{i,z}$$

$$B_y(\bm{r}) = \frac{\mu_0}{4\pi}\sum_i \frac{1}{|\bm{r}-\bm{r}_i|^5}\left(3(r_x - r_{i,x})(r_y - r_{i,y}) - |\bm{r}-\bm{r}_i|^2\right)m_{i,x}$$
$$+ \frac{1}{|\bm{r}-\bm{r}_i|^5}\left(3(r_y - r_{i,y})^2 - |\bm{r}-\bm{r}_i|^2\right)m_{i,y}$$
$$+ \frac{1}{|\bm{r}-\bm{r}_i|^5}\left(3(r_y - r_{i,y})(r_z - r_{i,z}) - |\bm{r}-\bm{r}_i|^2\right)m_{i,z}$$

$$B_z(\bm{r}) = \frac{\mu_0}{4\pi}\sum_i \frac{1}{|\bm{r}-\bm{r}_i|^5}\left(3(r_x - r_{i,x})(r_z - r_{i,z}) - |\bm{r}-\bm{r}_i|^2\right)m_{i,x}$$
$$+ \frac{1}{|\bm{r}-\bm{r}_i|^5}\left(3(r_y - r_{i,y})(r_z - r_{i,z}) - |\bm{r}-\bm{r}_i|^2\right)m_{i,y}$$
$$+ \frac{1}{|\bm{r}-\bm{r}_i|^5}\left(3(r_z - r_{i,z})^2 - |\bm{r}-\bm{r}_i|^2\right)m_{i,z}$$

This can be expressed as a linear superposition of dipole fields collected in a leadfield matrix $\bm{L_B}$

$$\bm{b} = \bm{L_B m},$$

where $b_k$ contains the vectorized flux density measurements at position (and direction) $k$ and $l_{kn}$ is the flux density due to a unit dipole located at position (and direction) $n$ evaluated at position $k$, as calculated by the relation above and $m_n$ is the dipole moment of dipole $n$. Estimation of the dipole moments amounts to solving this linear equation system for the vectorized dipole moments $\bm{m}$. Due to noise, high colinearity between the columns of $\bm{L}$ and a limited number of measurements available this problem is usually ill-posed. The solutions to these type of problems based on so-called minimum norm estimators are well-investigated in the literature dealing with source localization for EEG and MEG [9]. In the current setting we will investigate simple minimum norm estimates based on regularization with a shrinkage prior, which amounts to seeking the solutions to the problem that minimize the convex cost function *C(m)*:

$$C(\bm{m}) = \|\bm{b} - \bm{Lm}\|_F^2 + \alpha|\bm{\Gamma m}|^2,$$

where $\bm{\Gamma}$ is a projection matrix often used to impose smoothness or other desired properties of the solution, and $\alpha$ is a positive regularization parameter to be determined using cross-validation. In the current setting initial experiments indicated that it was not advantageous to impose smoothness of the solution via approaches such as low resolution electromagnetic tomography [10] and hence simply regularize using the L2 norm, e.g. $\bm{\Gamma} = \bm{I}$. Furthermore, we opted for split-half resampling, as a sufficiently dense sampling of the positions can easily be achieved in practice and reserving equal amounts of data for fitting and testing allows reproducibility of estimated parameters across splits to be assessed in a simple and convenient way [11]. For numerical solution of the Poisson equation governing the electric field, we require estimation of the magnetic vector potential **A** rather than **B**. In this setting, it is straightforward to obtain the **A** field using a similar linear leadfield equation

$$a = L_A m.$$

Due to the definition of **A** (eq. 2), it is defined up to addition of any curl-free field.

Varying the regularization parameter $\alpha$ results in different trade-offs between predictability (i.e. how well **m** estimated from one half-split of the measured **b** predicts the other half-split of the data) and reproducibility (i.e. how similar the **m** are that are estimated from the two half-splits of **b**). This trade-off is well-investigated for the type of inverse problems studied here and can visualized by predictability-reproducibility curves (figure 1c) [12]. Following standard practice, we optimized predictability and reproducibility jointly by minimizing the Manhattan distance to the point (R, P) = (1, 1). For that, we chose to measure both predictability and reproducibility as the coefficient of determination as it is naturally bounded between 0 and 1. In practice, we found that the joint optimization of predictability and reproducibility ensured smooth solutions for the calculated **B** and **A** also for non-measured positions very close to the coil surface. This improved in particular the accuracy of **A** in these regions that could otherwise suffer from local errors because **m** was optimized to fit the measured **B** rather than the *non-measurable* **A**.

## Simulated sparse data of a titled figure-8 coil

To demonstrate the general applicability of the approach, we determined the **B** fields of a simulated figure-8 coil with tilted wings using closed-form formulas (see Fig. 1d for details) and investigated the influence of noise on the reconstruction results by adding multiplicative Gaussian noise at various signal-to-noise ratios. Magnetic dipoles were placed on a regular grid inside the coil casing and the above method was applied to estimate their weights in order to best fit the simulated **B**. We tested two situations, first for **B** field data sampled in three planes above the coil (z=[33.7, 73.5, 113.2] mm; Fig. 1d illustrates the **B** field in an example plane), and second with one additional plane at the other side of the coil (z=-34.5 mm). Both situations were aimed to mimic measurement data that is sparse along the z-direction. For evaluation, we calculated the error of the fitted model as the difference of its **B** and **A** fields to the ground truth obtained by closed-form formulas. This was done on a new, more dense sampling constricted in a sphere with radius 75 mm and centered below the coil surface at a distance of 85 mm. These choices were aimed to roughly mimic an average head size and a minimal coil-cortex distance of 10 mm as worst case scenario.

## Reconstruction from physical measurements

A purpose-built setup (MagProbe, Skjøt Consulting Aps, Denmark; see [cite Maria] for details) with a three-axis Hall probe moved by stepper motors was used for measuring the magnetic flux density of a MagVenture MC-B70 coil in three dimensions. The sampled volume had a maximal area of 28x28 cm² parallel to the coil and extended ~24 cm in height. The closest point was 10 mm away from the coil. A part of the volume was V-shaped to align with the coil geometry. The sampling resolution was 8 mm in-plane and ~10 mm along the height. The field on both sides was acquired by turning the coil and repeating the measurements. The reconstruction method was applied to obtain the weights of magnetic dipoles placed on a regular grid with spacing of 5 mm covering positions inside the coil casing.

## Sparsification of the coil model

Pilot tests revealed that models consisting of around 1000 dipoles placed on a regular 5 mm grid inside the coil casing approximated the field very well even in regions close to the coil. We then aimed to explore to what extent sparsification of the dipole model is feasible. We first identified a solution with many dipoles (termed full model in the following) and then attempted to represent the same **A** field with fewer active dipoles using a similar minimization procedure as before, but this time aiming to match both the **A** and **B** fields of the dense model. It is important to note the amount of training data is not limited in this setting as it is provided via simulations of the fields of the full dipole model. The input data consisted of simulations of the **A** and **B** fields based on the original denser dipole model in 5000 random positions sampled uniformly in a sphere of radius 80 mm located with the center at (x,y,z)=(0,0,85). In addition, 300 random points were sampled in a plane -15 mm behind the minimum z-coordinate of the dipoles of the full model. The extent of the plane in x- and y-direction was given by the coil casing.

In order to obtain approximate equal contributions of the **A** and **B** fields during sparsification of the dipole model, the simulated data and leadfields were normalized such that the linear problem to be solved was

$$\begin{bmatrix} a_{\text{full}}/\|a_{\text{full}}\| \\ b_{\text{full}}/\|b_{\text{full}}\| \end{bmatrix} = \begin{bmatrix} L_A/\|a_{\text{full}}\| \\ L_B/\|b_{\text{full}}\| \end{bmatrix} m_{\text{sparse}} \,,$$

where $a_{\text{full}}$ and $b_{\text{full}}$ are the vectorized **A** and **B** fields simulated with the full dipole model, and $\|a_{\text{full}}\|$ and $\|b_{\text{full}}\|$ are the corresponding L2 norms. Leadfields $L_A$ and $L_B$ are for the **A** and **B** fields of the sparse model, and vector $m_{\text{sparse}}$ contains the dipole moments of the sparse dipole expansions. We then used the orthogonal matching pursuit algorithm [13] with an efficient implementation via the K-SVD algorithm [14] to provide an approximation of the full L0 regularization path of $m_{\text{sparse}}$ in the equation above. We sampled the dipole positions for the sparse solution in a denser grid (with 3 mm spacing) within the coil casing. However, in practice we found that the final solution was quite insensitive to the choice of dipole positions.

## Results

Our approach successfully recovers the **A** and **B** fields of the simulated non-flat figure-8 coil even when **B** field data of only in a few planes was used for fitting (Fig. 1a). The average residual error falls below 0.01 once the SNR simulated ground-truth **B** field exceeds 10 dB. This SNR level is far lower than the SNR of the real **B** fields measured with our setup (where the between split residual indicates an effective SNR of 31 dB), ensuring high reconstruction accuracy for real data. However, one plane on the backside of the coil is required to obtain an accurate estimation of the **A** field (see results for 4 vs. 3 planes in Fig. 1a). Of note, this plane is still required when the **B** field is measured densely in the region in front of the coil (data not shown).

As our approach uses a dipole expansion of the coil fields, the residual increases quickly for positions very close to the coil (Fig. 1b shows results for a SNR of 20 as example). However, for 4 "measurement" planes, the **A** field that is used as input to the electric field simulations is sufficiently accurate already a few millimeters away from the coil (Fig. 1d shows the corresponding dipole model).

The reconstruction from measured data of a MC-B70 coil works similarly well as for the simulated test case, as evident from the predictability-reproducibility curves in Figure 1c. The corresponding **B** and **A** fields are visualized in Figs. 2a&b, and the resulting dipole model is depicted in Fig. 2c ("full" model).

When sparsifying the dipole model, a natural tradeoff occurs between the number of dipoles used to represent the coil and the minimal distance to the coil surface at which the dipole field approximation is still adequate (Fig. 2d). Here, we evaluated the relative error for a coil to cortex distance, *x*, as the relative error on a half-sphere located 85 mm below the coil and with a radius of 85-*x* mm. Ensuring a relative error of the **A** field of less than 0.01 for a coil to cortex distance of 5 mm or more resulted in 74 dipoles as visualized in figure 2c ("sparse" model).

## Discussion

Our new method successfully reconstructs the **A** and **B** fields of TMS coils from measurements of their magnetic flux density, as demonstrated for both simulated ground truth and real measurement data. In contrast to our previous approach [5], it works for arbitrary winding geometries rather than only for flat coils and gives accurate results also for sparse input data. Here, we exemplarily tested sparsification of the input data along the z-axis, and we supply example scripts and data online that can be used to further pre-plan and optimize sparse sampling strategies that are adapted to the constraints of the specific measurement equipment.

In addition, our method allows for a post-hoc sparsification of the coil model with a flexible tradeoff between number of dipoles and the accuracy of the fields at a desired minimal distance. For the tested figure-8 coil, a model with 74 dipoles already gave a good accuracy for the **A** field at 5 mm distance, and a further reduction seems feasible when the aim is to guarantee accurate fields in the brain. While fast multipole methods [15,16] can be used to obtain the field from many sources in $\mathcal{O}(N)$ complexity, sparse dipole representation are still attractive in applications where computational efficiency is of key importance, such as updating electric field simulations in real-time [17] or for grid-search to obtain an optimal coil position [18].

Dipole expansions are attractive as they provide a compact representation and good approximation of fields distant from the source region. While the approximation will deteriorate quickly very close to the coil, this seems unproblematic for our application. The accuracy of the coil model is limited by the quality of the B field data, and hence requires a measurement setup which provides both accurate probe positions and B field measurement. Interestingly, information about the magnetic field beyond the reconstructed region was required to ensure adequate reconstruction of the **A** field. However, this information can be quite sparse as exemplified by adding one plane of data behind the coil in the current application. In addition to measured magnetic flux density data, the method requires the specification of a volume in which the dipoles can be placed, but no specific knowledge of the geometry of the coil windings. In practice, limiting the dipoles to a volume that approximates the coil casing worked well for us.

The code and usage examples are available at github.com/simnibs/dipole-fit under the GNU General Public License Version 3 to facilitate the adoption of our method.

## Acknowledgements

This study was supported by the Innovation Fund Denmark (grant 7038-00163B), the Lundbeck foundation (grant R313-2019-622) and the NIH (Grant No. 1RF1MH117428-01A1).

## Declaration of interest statement

Maria Drakaki was employed at MagVenture A/S (Farum, Denmark). The other authors report no conflicts of interests.

# Figures

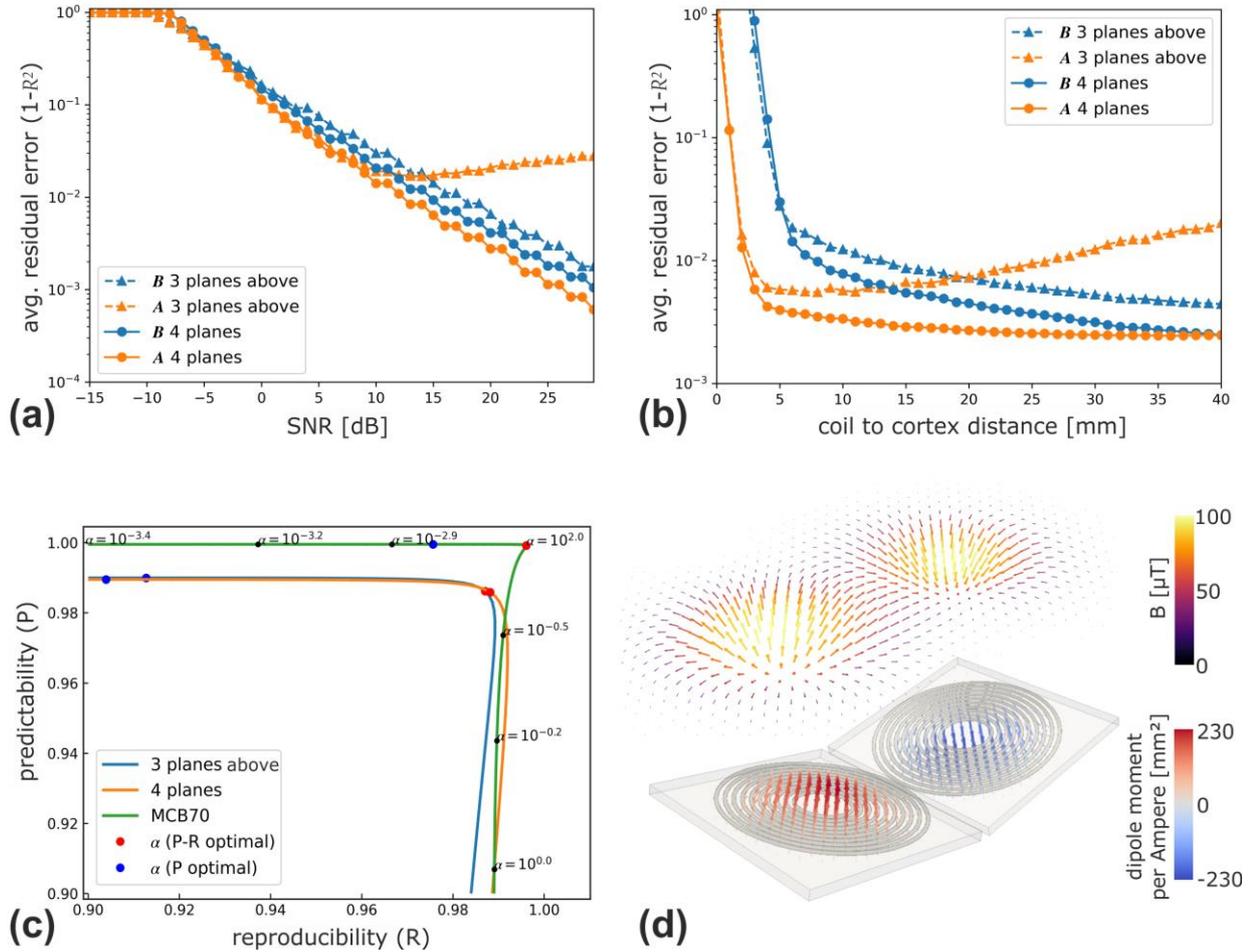

**Figure 1 (a)** Fitting of magnetic dipole models to simulated magnetic flux density data used as ground truth for varying signal-to-noise ratios (SNR). The magnetic flux density ***B*** and the corresponding magnetic vector potential ***A*** were obtained for the simulated figure-8 coil shown in subfigure d, using analytical expressions [19] for a current flow of 1 A. The signal-to-noise ratio (SNR) of ***B*** was controlled by adding multiplicative Gaussian i.i.d noise and the resulting data was provided as input to the model fitting. The data was sampled either only from 3 planes above the coil (at distances $z$=33.7, $z$=73.5 and $z$=113.2 mm) or additionally from a 4$^{th}$ plane placed below the coil ($z$=-34.5 mm). The plot shows the relative sum of squares error for the reconstructed ***B*** and ***A*** fields evaluated in a sphere centered 85 mm below the coil surface (center at (x,y,z)=(0,0,85) mm) and having a radius of 75 mm. While the fit of the reconstructed ***B*** improves continuously with increasing SNR, the fit of the reconstructed ***A*** levels off at higher SNRs (3 planes above) unless information about the magnetic flux density below the coil is provided as well (4 planes). **(b)** Error as function of the coil to cortex distance to the coil for a SNR of 20. The latter is a conservative estimate of the lower bound of the SNR achieved in the real measurements. This relative error is measured on the surfaces of concentric half-spheres centered 85 mm below the coil surface (center at (x,y,z)=(0,0,85) mm). The radii of the half-spheres were varied to obtain the minimal distances to the origin stated on the x-axis. **(c)** Predictability-Reproducibility curves when varying the regularization parameter α. The blue line (3 planes above) and orange line (4 planes) correspond to fits to the simulated magnetic flux density of the figure-8 coil shown in subpanel d for a SNR of 20dB. The green line indicates the fitting results to magnetic flux density data of a MagVenture MC-B70 coil measured densely in a 3D volume around the coil. Predictability is measured as the average coefficient of determination (R) of the reconstructed *B* across splits. Likewise, reproducibility of the fitted dipole moments *m* was assessed as the coefficient of determination of the two half-splits realizations for *m*. The red points denote the optimum as determined by the minimum Euclidean distance to (1, 1). The blue points mark the positions of highest reproducibility.  **(d)** Visualization of the simulated figure-8 coil with titled wings (angle ±11.5°). Each wing consisted of 10 wire loops (diameters equally spaced between 27.0 mm to 97.0 mm). The fitted magnetic dipoles obtained for data sampled from 4 planes at an SNR of 20dB are shown as blue and red arrows. The dipole positions were predefined on a regular grid inside the coil casing and the dipole moments then fitted to best explain the sampled *B* data. The corresponding reconstructed *B* for a distance of *z*=23.8 mm is shown in the plane above.

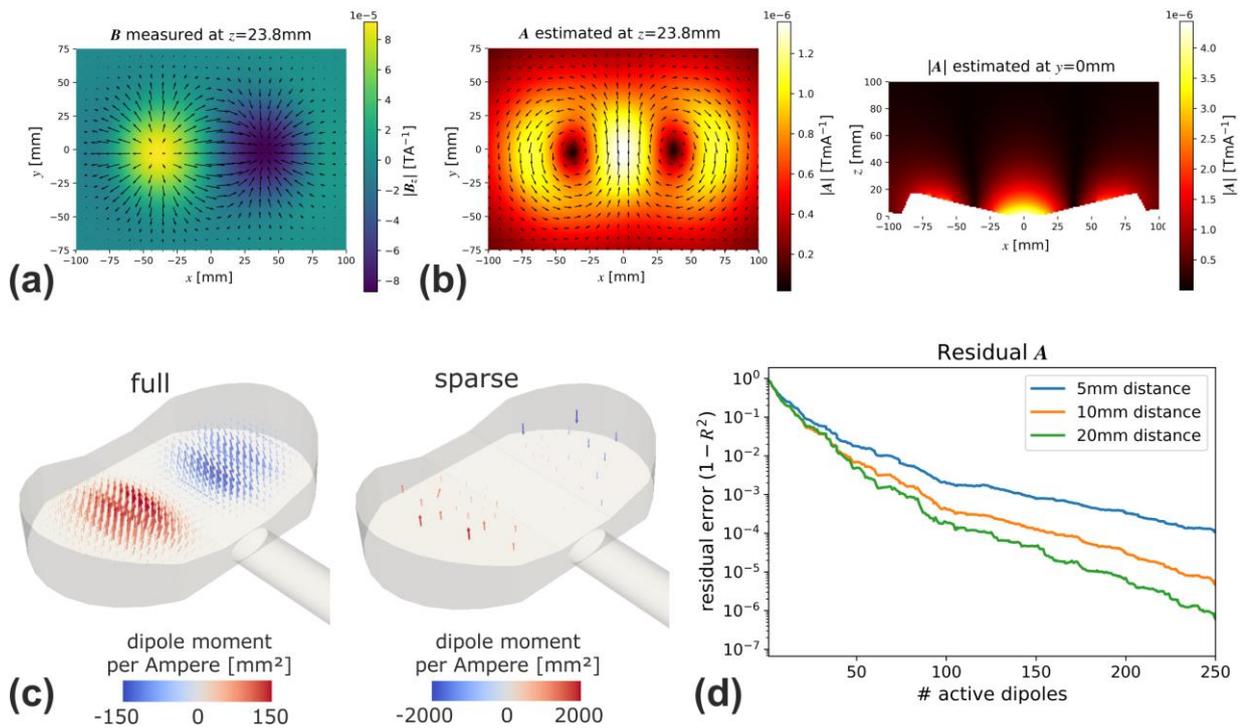

**Figure 2 (a)** Measured magnetic flux density of a MagVenture MC-B70 coil, exemplarily shown in a plane with a distance of 23.8 mm to the coil. The $B_z$ component is coded as color, and the general strength and direction of ***B*** is indicated by arrows. The reconstructed ***B*** is visually indistinguishable from the measured data and therefore not separately shown.  **(b)** Reconstructed ***A*** of the same coil, shown in one plane parallel to the coil surface (left panel) and one orthogonal plane (right panel). In the right panel, positions inside the coil casing are left blank.  **(c)** The fitted magnetic dipoles inside the casing of the MC-B70 coil are shown as blue and red arrows. The dipole positions were predefined on a regular grid inside the casing and the dipole moments then estimated to best fit the measured ***B***. The left panel corresponds to the full model with 1517 dipoles that was fitted without enforcing additional sparseness. The right panel corresponds to a sparse model with 74 dipoles that was fitted to obtain a relative residual error of maximally 1% (compared to the full model) for coil to cortex distances of 5 mm or more.  **(d)** Dependence of the accuracy of the reconstructed ***A*** compared to the full model on the number of dipoles for three different coil to cortex distances (blue 5 mm, orange 10 mm and green 20 mm). The former corresponds to a coil model that gives accurate electric field estimates already at positions very close to the coil (e.g. inside the scalp), while the latter corresponds to a coil model with good accuracy for positions in the brain.

# References

[1]   Thielscher A, Kammer T. Electric field properties of two commercial figure-8 coils in TMS: Calculation of focality and efficiency. Clin Neurophysiol 2004;115. doi:10.1016/j.clinph.2004.02.019.

[2]   Deng ZD, Lisanby SH, Peterchev A V. Electric field depth-focality tradeoff in transcranial magnetic stimulation: Simulation comparison of 50 coil designs. Brain Stimul 2013;6:1–13. doi:DOI 10.1016/j.brs.2012.02.005.

[3]   Petrov PI, Mandija S, Sommer IEC, Van Den Berg CAT, Neggers SFW. How much detail is needed in modeling a transcranial magnetic stimulation figure-8 coil: Measurements and brain simulations. PLoS One 2017;12:1–20. doi:10.1371/journal.pone.0178952.

[4]   Bohning DE, Pecheny AP, Epstein CM, Speer AM, Vincent DJ, Dannels W, et al. Mapping transcranial magnetic stimulation (TMS) fields in vivo with MRI. Neuroreport 1997;8:2535–8. doi:10.1097/00001756-199707280-00023.

[5]   Madsen KH, Ewald L, Siebner HR, Thielscher A. Transcranial Magnetic Stimulation: An Automated Procedure to Obtain Coil-specific Models for Field Calculations. Brain Stimul 2015;8:1205–8. doi:10.1016/j.brs.2015.07.035.

[6]   Nieminen JO, Koponen LM, Ilmoniemi RJ. Brain Stimulation Experimental Characterization of the Electric Field Distribution Induced by TMS Devices. Brain Stimul 2015;8:582–9. doi:10.1016/j.brs.2015.01.004.

[7]   Thielscher A, Opitz A, Windhoff M. Impact of the gyral geometry on the electric field induced by transcranial magnetic stimulation. Neuroimage 2011;54. doi:10.1016/j.neuroimage.2010.07.061.

[8]   Nummenmaa A, Stenroos M, Ilmoniemi RJ, Okada YC, Hamalainen MS, Raij T. Comparison of spherical and realistically shaped boundary element head models for transcranial magnetic stimulation navigation. Clin Neurophysiol 2013;124:1995–2007. doi:DOI 10.1016/j.clinph.2013.04.019.

[9]   Hämäläinen MS, Ilmoniemi RJ. Interpreting magnetic fields of the brain: minimum norm estimates. Med Biol Eng Comput 1994;32:35–42. doi:10.1007/BF02512476.

[10]  Pascual-Marqui RD, Michel CM, Lehmann D. Low resolution electromagnetic tomography: a new method for localizing electrical activity in the brain. Int J Psychophysiol 1994;18:49–65. doi:10.1016/0167-8760(84)90014-X.

[11]  Strother SC, Anderson J, Hansen LK, Kjems U, Kustra R, Sidtis J, et al. The quantitative evaluation of functional neuroimaging experiments: the NPAIRS data analysis framework. Neuroimage 2002;15:747–71. doi:10.1006/nimg.2001.1034.

[12]  Rasmussen PM, Hansen LK, Madsen KH, Churchill NW, Strother SC. Model sparsity and brain pattern interpretation of classification models in neuroimaging. Pattern Recognit 2012;45:2085–100. doi:10.1016/j.patcog.2011.09.011.

[13]  Mallat S, Zhang Z. Matching pursuits with time-frequency dictionaries. IEEE Trans Signal Process

1993;41:3397–415.

[14] Rubinstein R, Zibulevsky M, Elad M. Efficient Implementation of the K-SVD Algorithm using Batch Orthogonal Matching Pursuit, 2008.

[15] Cheng H, Greengard L, Rokhlin V. A Fast Adaptive Multipole Algorithm in Three Dimensions. J Comput Phys 1999;155:468–98. doi:10.1006/jcph.1999.6355.

[16] Rokhlin V. Rapid solution of integral equations of classical potential theory. J Comput Phys 1985;60:187–207. doi:10.1016/0021-9991(85)90002-6.

[17] Stenroos M, Koponen LM. Real-time computation of the TMS-induced electric field in a realistic head model. Neuroimage 2019;203:116159. doi:10.1016/j.neuroimage.2019.116159.

[18] Weise K, Numssen O, Thielscher A, Hartwigsen G, Kn TR. A novel approach to localize cortical TMS effects. Neuroimage 2020;209:116486. doi:10.1016/j.neuroimage.2019.116486.

[19] Simpson JC, Lane JE, Immer CD, Youngquist RC, Steinrock T. Simple Analytic Expressions for the Magnetic Field of a Circular Current Loop. 2001.